\documentclass[aps,prl,superscriptaddress,showpacs,twocolumn]{revtex4}

\usepackage{graphicx}
\usepackage{bm}

\begin{document}


\title {Generalized Spectral Signatures of Electron
  Fractionalization in Quasi-One and -Two Dimensional Molybdenum
Bronzes and Superconducting Cuprates}

\author {G.-H. Gweon$^\dagger$}
\author {J. W. Allen}

\affiliation {Randall Laboratory of Physics, University of Michigan, 500
  E. University, Ann Arbor, MI 48109}

\author {J. D. Denlinger} 

\affiliation {Advanced Light Source, Lawrence Berkeley National
  Laboratory, Berkeley, CA 94720}

\date {\today}

\begin {abstract}
  
  We establish the quasi-one-dimensional Li purple bronze 
as a photoemission paradigm of Luttinger liquid behavior.  We
also show that generalized signatures of electron fractionalization 
are present in the angle resolved photoemission 
spectra for quasi-two-dimensional purple bronzes and certain cuprates.
An important component of our analysis for the quasi-two-dimensional
systems is the proposal of a ``melted holon'' scenario for the 
\textbf{k}-independent background that accompanies but does not interact 
with the peaks that disperse to define the Fermi surface. 

\end {abstract}

\pacs {71.10.Pm, 71.20.-b, 79.60.-i}
\maketitle

\section{Introduction}

In the Landau Fermi-Liquid (FL) theory \cite{landau-FL} of interacting
electrons, low energy single-particle, i.e.\ electron addition and
removal, excitations behave like free electrons except for enhanced
mass and therefore are termed Landau ``quasi-particles.'' The electron in
this generalized sense retains its status as a fundamental particle.  One 
paradigm for non-FL (NFL) behavior is the electron fractionalization that occurs
in the so-called Luttinger liquid (LL), as defined
\cite{haldane-LL} by the phenomenological applicability of the Tomonaga-Luttinger (TL) model \cite{TLmodel-T,TLmodel-L} of
interacting electrons in one-dimension. In this new state, electron-like
single-particle eigenstates no longer exist, and an electron/hole
excitation propagates only as a continuum of collective density waves.  In
this sense, the electron is ``fractionalized.''
\cite{anderson-efrac,kivelson-efrac,orgad-efrac}  Two key 
features of LL fractionalization 
are power law behaviors of correlation functions, characterized by an
anomalous dimension ($\alpha$), and the complete separation of spin 
and charge degrees of freedom into density waves dispersing with 
different velocities $\text{v}_{\text{s}}$, $\text{v}_{\text{c}}$.
The Luther-Emery (LE) model \cite{LEmodel} differs from the TL model by an
additional interaction term that causes a gap in the spin excitation
spectrum. In this paper we focus on the single particle spectral 
function, which has characteristic features that are much different 
from that of a FL. This spectral function can be measured in 
angle resolved photoemission spectroscopy (ARPES).

In spite of many efforts on ARPES in quasi-low-d systems 
\cite{dardel-prl,grioni-voit,gweon-jesrp}, there remains
much skepticism as to the relevance of fractionalization to these 
spectra. First, the TL model is strictly one-dimensional. 
It is then a serious question whether actual quasi-1-d materials 
with their underlying higher dimensionality could display TL 
(or LE) behavior and beyond that, whether such behavior
could be found even in quasi-2-d materials such as the superconducting
cuprates, as has been proposed \cite{anderson-efrac,kivelson-efrac}
 in various scenarios.  Second,
the ARPES spectra of low dimensional materials generally 
differ in various ways from the simple model spectra.  
Nonetheless the experimental spectra display definite NFL attributes. 
Our stance here is that the simple models correctly show generic 
possibilities while lacking one or more of the elements needed for 
describing actual data.  Treating all the important elements, e.g.
multiple bands, the full Coulomb interaction, electron-phonon interactions,
and higher dimensional couplings, all on an equal footing, is beyond 
current theory. But we surmise that fractionalization may be 
the correct physics underlying the more complex reality, even 
for some materials with dimensionality greater than one.

\begin{figure}

  {\centering \resizebox*{0.95\columnwidth}{!}{\includegraphics{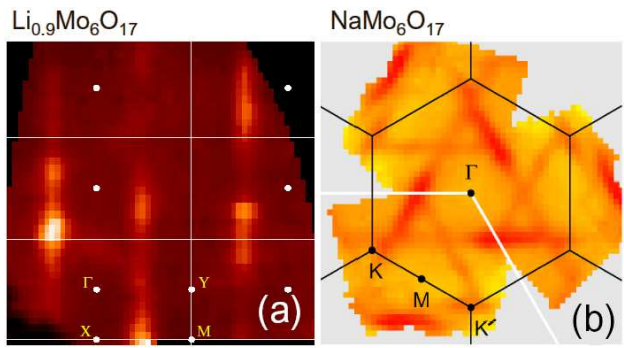}}
    \par}

\caption{\label{fig:bronzefs}
  ARPES FS maps for (a) quasi-1-d Li purple bronze 
  \cite{denlinger-Li-purple-prl} and (b) 
  quasi-2-d (hidden 1-d) Na purple bronze
  \cite{gweon-hidden-1d-FS}.}
\end{figure}

In this paper we adduce strong evidence supporting 
such a hypothesis.  We have obtained
over time \cite{denlinger-Li-purple-prl,gweon-prlComm,gweon-jesrp,gweon-physicaB} increasingly more detailed ARPES 
spectra of the quasi-1-d metal Li$_{0.9}$Mo$_6$O$_{17}$ 
(Li``purple bronze'').  The 1-d Fermi surface (FS) found 
in these studies \cite{denlinger-Li-purple-prl} is shown in 
panel (a) of Fig.\ \ref{fig:bronzefs}.  Here we show 
that, with the exception of one aspect that is nonetheless
of definite NFL character,  the dispersing lineshapes defining 
the ARPES FS are well described by finite temperature 
TL model theoretical spectra \cite{orgad-finite-T-TL-LE}.
This material is thus an LL ARPES paradigm.
We also introduce generalized signatures 
of electron fractionalization extracted from the TL model 
and related thinking.
Presentation of these signatures provides the basic 
organization of the paper and we 
show that they are present in our ARPES spectra
of several low dimensional bronzes 
\cite{schlenker-Mo-oxide-book}, including those of 
quasi-1-d K$_{0.3}$MoO$_3$ (the K ``blue bronze'') and 
of quasi-2-d NaMo$_6$O$_{17}$ (the Na ``purple bronze'').  The
Na purple bronze is significant here as a bridge material 
between 1-d and 2-d because of its ``hidden 1-d''
FS \cite{whangbo} arising from three weakly coupled 1-d chains 
that are mutually oriented at 120 degrees. We have
previously measured its FS \cite{gweon-hidden-1d-FS} by ARPES, 
as reproduced in panel (b) of Fig.\ \ref{fig:bronzefs} and
in this paper we show that the associated ARPES spectra display 
generalized fractionalization 
signatures that are also present in the ARPES spectra of the quasi-2-d 
superconducting cuprate Bi$_2$Sr$_2$CaCu$_2$O$_{8+\delta}$ 
(Bi2212) \cite{kaminski-prl,dessau-Bi2212-prl93}. An 
important component of the analysis is to identify the 
effect of charge disorder on the LL spectra.


We are concerned here only with normal state line shapes above phase
transitions to ordered states, e.g. charge density wave (CDW), 
spin density wave (SDW), or superconductivity (SC).  Whereas the ordered
state must entail higher dimensionality the normal state is
expected to display quasi-1-d fractionalization behavior \cite{Carlson}.  
In this connection we note that models with electron-phonon 
interactions but no explicit electron-electron interactions 
show above $T_{CDW}$ a CDW-fluctuation-induced pseudo-gapped NFL metallic state
\cite{mckenzie-SRO-non-FL}, which, we postulate, might also admit of a
fractionalization description. The phase transitions of our materials 
occur at 24 K (Li purple bronze: order unknown but
\textit{not} CDW/SDW \cite{denlinger-Li-purple-prl,gweon-jesrp}), 
180 K (K blue bronze: CDW), 80 K (Na purple bronze: CDW), and 90 K (Bi2212:
SC).  X-ray scattering \cite{pouget-Li-struct} due to CDW or SDW 
formation has never been found in the Li purple bronze 
and its optical conductivity \cite{degiorgi-purple-optics} 
shows no gap down to 1 meV.  It is however a SC
below 1.9 K \cite{greenblatt}.  For the K blue and 
Na purple bronzes CDW fluctuation
effects above $T_{CDW}$ have been identified in X-ray diffraction up to
room temperature \cite{schlenker-Mo-oxide-book}.

Throughout this paper,
a single but typical ARPES data set is discussed for each material, plus
data (Figs.\ \ref{fig:aipes}(b-d)) taken for bronzes with an
angle-integrated VG ESCALAB II spectrometer.  The ARPES data were
taken along 1-d chain axes for bronzes, and along the $(\pi,\pi)$
direction, i.e.\ the diagonal of the 2-d Brillouin zone (BZ) for Bi2212
\cite{kaminski-prl}. All of the data presented are already published
\cite{gweon-jesrp,gweon-physicaB}, with two exceptions
(Figs.\ \ref{fig:aipes}(b) and \ref{fig:ar}(d) \cite{ssrl-data}). In the
following, we give values of the experimental energy resolution $\Delta E$
(in FWHM) and the full angular width $\Delta \theta$, where it
becomes relevant to do so.

\section{Generalized Fractionalization Signatures}

\textbf{1. No Fermi Edge in Angle-Summed Spectrum}\ \ Electron (or hole)
fractionalization in the LL leads to a strong and surprisingly
counter-intuitive suppression of the {\bf k}-summed spectral function
$\rho(\omega)$ at the chemical potential ($\mu$), {\em vanishing} as a
power law $\vert \omega - \mu \vert ^\alpha$ at $T=0$.  This is contrasted
with the Fermi edge (FE) line shape of an FL, given by the Fermi-Dirac
function.  $\rho(\omega)$, broadened by $\Delta E$, is measured by the
angle-summed ARPES and experimental data hinting at non-zero $\alpha$ have
been known for some time \cite{dardel-prl}.

\begin{figure}

  {\centering \resizebox*{0.95\columnwidth}{!}{\includegraphics{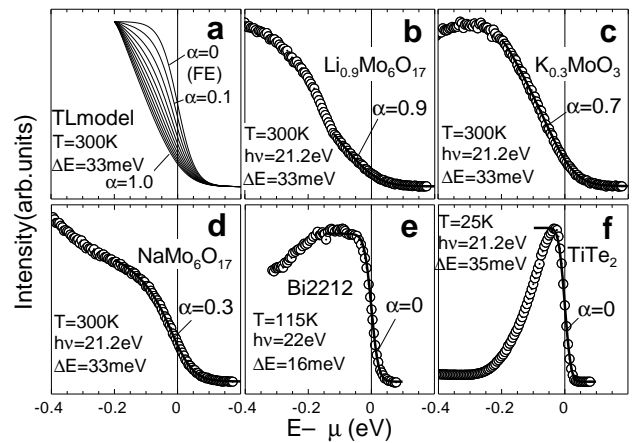}}
    \par}

\caption{\label{fig:aipes}
  Theoretical \cite{orgad-finite-T-TL-LE} and experimental angle-summed
  spectral functions.  Best-fit curves from panel (a), with modified
  $\Delta E$ and $T$ values where appropriate, are shown as lines.}
\end{figure}

We compare various angle-summed spectra in Fig.\ \ref{fig:aipes} with
theoretical line shapes.  As an experimental FL reference
 \cite{claessen-prl-TiTe2,claessen-prb-TiTe2,perfetti-TiTe2}, we include the
angle-sum of the TiTe$_2$ ARPES data \cite{claessen-prl-TiTe2}.  For
comparison to the 1-d TL lineshapes, our angle integrated spectra of the
quasi-1-d and hidden 1-d bronzes are effectively 1-d sums and for Bi2212
we sum ARPES spectra \cite{kaminski-prl} along a line normal to the FS.
One immediately notices a qualitative difference between the smooth
onsets at $\mu$ of the quasi-1-d metals, definitely non-FL, and the more abrupt
edges of the quasi-2-d metals, all resembling a FE.  However, fits to the
theoretical line shapes of panel (a) give $\alpha$ values that require
more refined thinking: 0.9 (Li purple), 0.7 (K blue), 0.3 (Na purple), and
0 (Bi2212, TiTe$_2$).  Two new findings, enabled by the use here 
of the recent finite $T$
spectral function theory \cite{orgad-finite-T-TL-LE} of the TL model, are
significant for the discussion below: a {\em finite} $\alpha$ for the {\em
  quasi-2-d} Na purple bronze and an $\alpha$ value increased from that
(0.6) obtained \cite{denlinger-Li-purple-prl} previously by
applying a T=0 theory to the same data
for the Li purple bronze.  The FE line shape for Bi2212 is in
apparent contradiction to its non-FL properties
\cite{anderson-efrac,kivelson-efrac}.  Along this particular 
line of {\bf k}-space
\cite{alpha-anisotropy} it is possible that the $\alpha$ value is so small
that the spectrum is essentially the same as an FE for the given $T$ and
$\Delta E$.  But the second signature along
this line, presented below, yields a moderately large $\alpha$ value,
making it also possible that instead the energy range of the power law
behavior is so small that it is completely masked by $T$ and $\Delta E$.
Indeed theoretical work shows that such a small energy range, on the order
of 10 meV, can occur for the 1-d Hubbard model \cite{penc-1d-Hubbard}.  A
Hubbard model is likely for cuprates but not for the Mo bronzes whose 4d
orbitals are less localized.

Some comments are in order for the fits of Fig.\ \ref{fig:aipes}.  First,
the TL model used here could be replaced by any theory which yields a
power law spectrum at $T=0$ and obeys quantum critical scaling at finite
$T$.  Second, the smallness ($\Delta \approx 20$ meV) of the spin
pseudo-gap \cite{johnston}, which implicates the LE model for the blue
bronze \cite{grioni-voit}, justifies its neglect here.  More specifically,
in this paper we limit ourselves to the case max$(T,\Delta E) >
\Delta(T)$, where $\Delta(T)$ is any pseudo-gap.  Third, the TL model
assumes a constant one-electron density of states (DOS).  Thus the
$\alpha$ values obtained by the fit potentially have some un-quantified
errors and so for the Na purple bronze we have verified explicitly that
an FL line shape with a sloped DOS does not produce a good fit.  Fourth,
the energy range of the fit is determined a posteriori by the fit, and is
marked by the energy range of the lines, except for TiTe$_2$, for which
only the FE width is fit because its unusually small band width
\cite{claessen-prl-TiTe2} causes the strong drop of intensity as soon as
the binding energy becomes bigger than the FE width.  We note, in
particular, that the fit range for the Li purple bronze is binding energy
$< 0.12$ eV, at which point two non-$\mu$-crossing bands
\cite{gweon-jesrp,gweon-physicaB} start to contribute and give rise to the
break observed in the line shape of panel (b).

\begin{figure}

  {\centering \resizebox*{0.95\columnwidth}{!}{\includegraphics{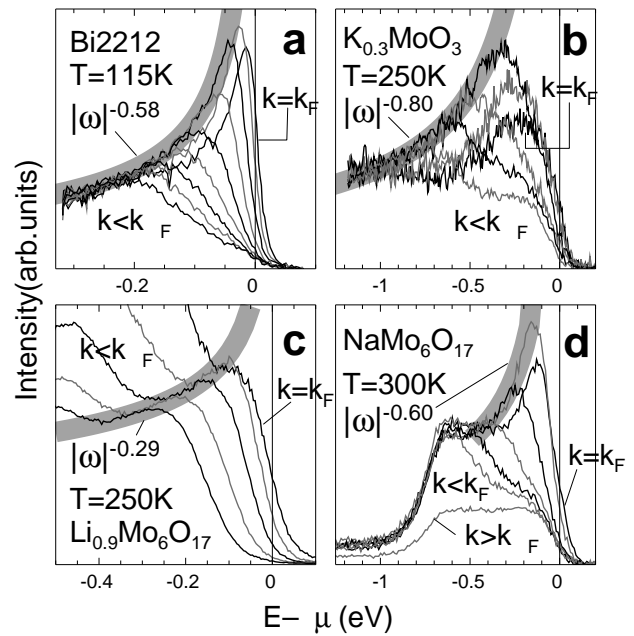}}
    \par}

\caption{\label{fig:ar}
  Anderson-Ren plots for (a) the Bi2212 data taken along $\Gamma$--Y
  \cite{kaminski-prl}, (b) the quasi-1-d band for the K blue bronze
  \cite{gweon-jesrp}, (c) the quasi-1-d band for the Li purple bronze
  \cite{gweon-physicaB}, and (d) the hidden-1-d band crossing $\mu$ for
  the Na purple bronze.  The data in (d) are shown again in Fig.\ 
  \ref{fig:napb2212} (a) as a stack plot with momentum value labels.  For
  these plots, the intensity scale for individual curve was varied
  as a free parameter.}
\end{figure}

\textbf{2. Anderson-Ren Line Shape}\ \ This is another signature that we
associate with $\alpha$.  Anderson and Ren (AR) proposed
\cite{anderson-ren-line-shape} an empirical visualization scheme for the
Bi2212 ARPES line shapes along the diagonal \cite{alpha-anisotropy} of the
2-d BZ\@.  In this view, the line shape is given by a common power law
tail relative to $\mu$, with exponent $\alpha - 1$, for binding energies
greater than the peak position, and a straight line fall to $\mu$ for
energies less than the peak position.  Fig.\ \ref{fig:ar} shows the
dispersing line shapes of Bi2212 and of the bronzes, plotted 
with intensity scaling to show the
remarkable result that the AR line shape is observed in \textit{all} of
them.  In contrast, we find that the FL ARPES line shapes for the Ti 3d
band of TiTe$_2$ \cite{claessen-prl-TiTe2} and also for a Mo surface state
\cite{valla-Mo} do not follow the AR line shape.

\begin{figure}

  {\centering \resizebox*{0.95\columnwidth}{!}{\includegraphics{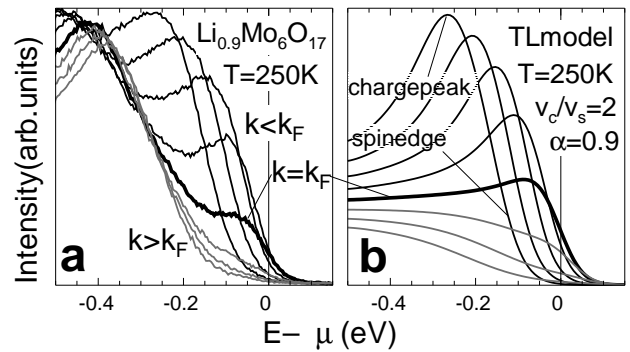}}
    \par}

\caption{\label{fig:li}
  High resolution angle resolved photoemission data for the Li purple
  bronze \cite{gweon-physicaB} taken along $\Gamma$-Y (a) and TL model
  \cite{orgad-finite-T-TL-LE} simulation (b).  The data show a single band
  crossing, due to the suppression of the other $\mu$-crossing band in
  this geometry \cite{gweon-physicaB}.  In the data, the peaks with energy
  $<$ -0.3 eV arise from a non-$\mu$-crossing band
  \cite{gweon-jesrp}, which is excluded from the theoretical
  simulation.  As the bottom curve (corresponding to \textbf{k} $\gg$
  \textbf{k}$_{\text{F}})$ of panel (a) shows, the line shape of the
  non-$\mu$-crossing band is a well confined peak, so the line shape of
  the $\mu$-crossing band can be observed essentially unhindered in an
  extended energy range, e.g.\ energies greater than its peak energy minus
  $\approx$0.1 eV for \textbf{k} $\le$ \textbf{k}$_{\text{F}}$.
  Experimental conditions ($T=250$K, $\Delta E=49$ meV, $\Delta \theta =
  0.36^o$) are fully included in the simulation.}
\end{figure}

We examine next the relationship between the exponents from the AR line
shape and from the $\mu$ onset in angle-summed spectra.  The plots in
Fig.\ \ref{fig:ar} show that the AR line shape does not hold when
\textbf{k} approaches \textbf{k}$_{\text{F}}$.  Therefore, the AR line
shape, as identified here, is a high energy and high momentum behavior.
Nonetheless, taking the AR connection to $\alpha$, we get $\alpha$ values
of 0.4 (Bi2212), 0.2 (K blue), 0.7 (Li purple), and 0.4 (Na purple).
These values show reasonable agreement with those found in Fig.\ 
\ref{fig:aipes} for the Li and Na purple bronzes, but not for the K blue
bronze and Bi2212.  For the K blue bronze this finding could signal that
its spin pseudo-gap breaks the AR connection to $\alpha$.  For Bi2212 we
infer that the $\alpha=0$ result in Fig.\ \ref{fig:aipes} reflects an
unobservably small energy range for power law behavior near $\mu$, as
discussed above.  The Li purple
bronze ARPES data that we reported initially \cite{denlinger-Li-purple-prl}
give an exponent of -0.2 (hence $\alpha=0.8$) in reasonable agreement with
the current data.  We also verified that the tail exponent reported in
Fig.\ \ref{fig:ar} is modified only slightly to -0.35, if we subtract
out approximately the contribution from higher energy non-$\mu$-crossing 
bands (see Fig.\ \ref{fig:li} caption).  In short,
our analysis distinguishes large $\alpha$ $(> 0.5)$ correlations found in
the Li purple bronze and small $\alpha$ $(< 0.5)$ correlations found in
the Na purple bronze and Bi2212, important for discussing the next
signature.

Before proceeding, we mention that there is no known microscopic
derivation for the AR line shape, yet \cite{AR 
disordered}.  It is easy to see that the 
LL fails because it produces
line shapes for which the origin of the approximate power law is the moving
position of the dispersive peak rather than the fixed $\mu$.  This
observation is intriguing, particularly in light of the otherwise
excellent LL description of the Li purple bronze, as we will see now.

\begin{figure}

  {\centering \resizebox*{0.95\columnwidth}{!}{\includegraphics{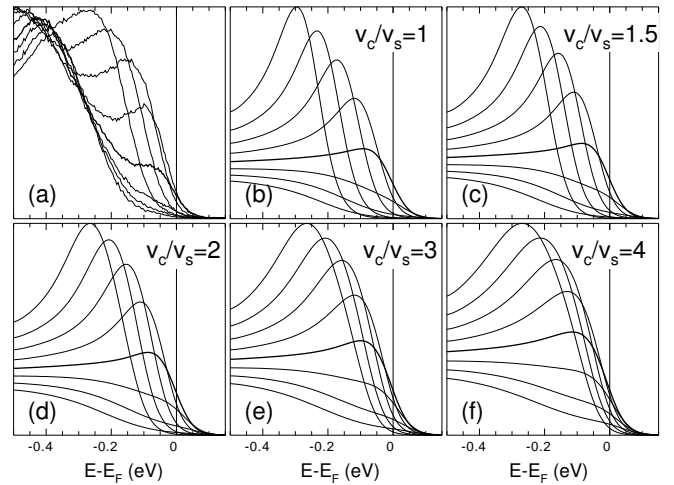}}
    \par}

\caption{\label{fig:tlvsvc}
  TL model spectra from theory of Ref.\ \cite{orgad-finite-T-TL-LE} 
 in panels (b) through (f) compared to Li purple 
 bronze ARPES data of panel (a) and Fig.\ \ref{fig:li} (a)
 to show sensitivity of TL description of data to choice of ratio of 
 velocities of holon peaks and spinon edges.  Holon peak dispersion is
 held constant and matched to experimental peak dispersion for ease of
 comparison of spinon edge dispersions.}
\end{figure}

\begin{figure}

  {\centering \resizebox*{0.95\columnwidth}{!}{\includegraphics{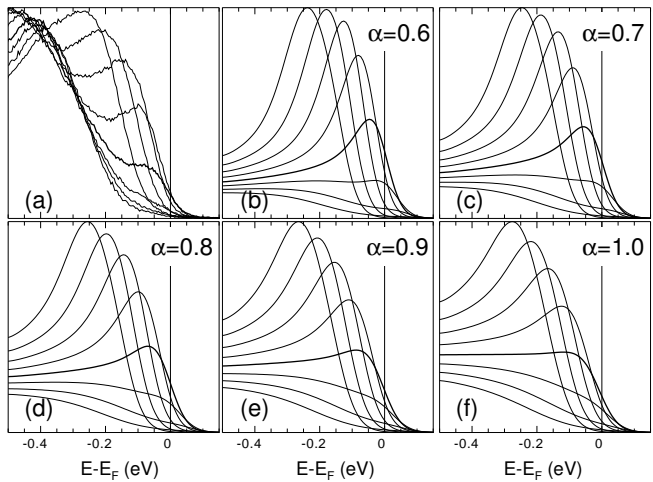}}
    \par}

\caption{\label{fig:tlalpha}
  TL model spectra from theory of Ref.\ \cite{orgad-finite-T-TL-LE} 
 with $\text{v}_{\text{c}}$/$\text{v}_{\text{s}}$
 value of 2 and various $\alpha$ in panels (b) through 
(f) compared to Li purple bronze ARPES data of panel (a) and 
 Fig.\ \ref{fig:li} (a) to show sensitivity of TL description 
 of data, with regard to the weight at $\mu$ and the broadness of the peak
 at $\mu$ crossing, to choice of $\alpha$.}
\end{figure}

\textbf{3. Two or More Objects in ARPES}\ \ In the LL lineshape, two features
disperse with velocities $\text{v}_{\text{c}}$, $\text{v}_{\text{s}}$, in
contrast to the single quasi-particle peak of the FL.  Depending on the
magnitude of $\alpha$, the low binding energy spin feature is either a
peak ($\alpha < 0.5$) or an edge ($\alpha > 0.5$)
\cite{meden-voit-LL-line-shape}.  As we discussed and demonstrated before
\cite{denlinger-Li-purple-prl}, by reason of its ideally 1-d FS, 
linear band dispersion to at least 300 meV below $\mu$, low 
transition temperature (24 K), and lack of a single particle gap opening
at least down to 1 meV in the low-T phase, the Li purple bronze is a unique
candidate for a simple LL description, an $\alpha =0.9$ TL in
particular. Fig.\ \ref{fig:li} shows our newest data set
\cite{gweon-physicaB} compared with the new finite
$T$ theory \cite{orgad-finite-T-TL-LE}.  The same $\alpha$ value, 0.9,
that best simulates the amount of the weight at the $\mu$ crossing relative 
to the peak height
also agrees nicely with the value obtained in
Fig.\ \ref{fig:aipes}.  The $\text{v}_{\text{c}}$/$\text{v}_{\text{s}}$
value of 2 (and $\hbar \text{v}_{\text{c}}$=4eV{\AA}) is used as in Ref.\ 
[\onlinecite{gweon-physicaB}], where we already noted that the improved
angle resolution $\Delta \theta$ relative to that of our earliest work
\cite{denlinger-Li-purple-prl} gives better resolution of the spin edge
and leads to a $\text{v}_{\text{c}}$/$\text{v}_{\text{s}}$ value changed
from our early value of 5.  Overall, there is excellent agreement between
the improved experiment and the improved theory.  Fig.\ \ref{fig:tlvsvc} 
shows that the quality of the 
agreement is definitely sensitive to the choice 
of $\text{v}_{\text{c}}$/$\text{v}_{\text{s}}$ and 
Fig.\ \ref{fig:tlalpha} shows the same for the choice of
$\alpha$.  Other features such as
the general decrease of intensity as the peak approaches $\mu$ and the $\mu$
weight retraction after the peak has crossed $\mu$ are also reproduced with
internal consistency by the theory.  The same is true, of course, for the
$\mu$ onset of the angle summed spectrum, and in this context 
we emphasize again the enormous difference between the angle summed 
spectra of the Fermi liquid reference material TiTe$_2$ and of the 
Li purple bronze.  The excellent agreement with the TL
theory makes the Li purple bronze presently unique (apart 
from the AR tails which are nonetheless of clear NFL character)
as an ARPES example of the TL line shape.

\begin{figure}

  {\centering \resizebox*{0.95\columnwidth}{!}{\includegraphics{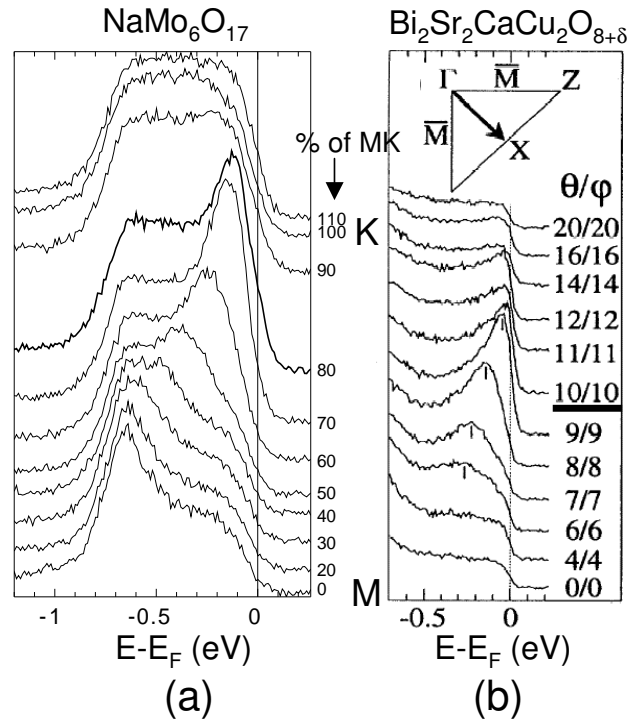}}
    \par}

\caption{\label{fig:napb2212}
  ARPES lineshapes (a) along M-K for the
 quasi-2-d Na purple bronze \cite{ssrl-data} compared to those
 (b) along $\Gamma$-X for the 
 Bi2212 \cite{dessau-Bi2212-prl93}.}
\end{figure}

We generalize the idea of two dispersing objects in the TL line shape as
follows. As shown in panel (d) of Fig.\ \ref{fig:ar}
and panel (a) of Fig.\ \ref{fig:napb2212}, the line shapes of the 
2-d Na (or K) purple bronze \cite{gweon-jpcm} are surprisingly
complex considering the simplicity and good agreement with band
theory of the ARPES FS shown in Fig.\ \ref{fig:bronzefs}.  In particular,
the lineshapes show two \textit{independent} components, 
\textit{both} the dispersing peaks that give the well
defined FS \textit{and} an equally large amount of \textbf{k}-independent
weight seen by itself in the \textbf{k} $>$ \textbf{k}$_{\text{F}}$
spectra. Because this weight is perfectly confined
within the bandwidth it does not have an extrinsic origin, e.g.\ the
inelastic scattering of photoelectrons. Charge disorder scattering giving
\textbf{k}-loss is likely, due to the well-known tendency for alkali atom
deficiencies in the purple bronzes \cite{schlenker-Mo-oxide-book}, 
but in a FL picture it is impossible to understand the \textit{fractionalized}
\textbf{\textit{k}}\textit{-loss} seen here. The lack of apparent 
interaction between this weight and the dispersing peak, 
and its presence unreduced in amplitude after the peak crosses $\mu$, 
shows that it cannot be simply attributed to the incoherent 
part of a FL spectral function.

We propose electron
fractionalization, with badly scattered charge waves giving the
\textbf{k}-independent weight and spin waves, un-scattered because they do
not see the charge disorder, 
dispersing to define the FS just as they do
in the Li purple bronze. Here, the spin waves are peaks instead of edges
because $\alpha < 0.5$. Having now a clear example, augmented 
by other fractionalization signatures, confers much
plausibility on the original suggestion \cite{anderson-private} to us
of this ``melted holon'' picture for Bi2212 (and other metallic cuprate)
spectra. 
As seen in panel (b) of Fig.\ \ref{fig:napb2212}, these spectra also show, 
in addition to dispersing peaks, \textbf{k} independent 
weight \cite{dessau-Bi2212-prl93} that is remarkably similar to that 
of the Na purple bronze except that the overlap with oxygen
bands precludes knowing if it is also so neatly confined to the d-bandwidth.
It is significant that the blue bronze has neither charge disorder nor
\textbf{k}-independent weight.  The 1-d Li purple bronze has charge
disorder and again shows some \textbf{k}-independent weight \cite{gweon-jesrp}
although less than that of the Na purple bronze.  Since the Li purple bronze
spectra show both dispersing charge peaks and spin edges, this weight is
perhaps some small portion of the charge peak weight but could signal the
more intriguing possibility that fractionalization has occurred into
\textit{three} density excitations, a possibility which is actually known
in theory \cite{balents-coupled-LL}, and is plausible in this material
with two bands crossing $\mu$ together.


\begin{figure}

  {\centering \resizebox*{0.95\columnwidth}{!}{\includegraphics{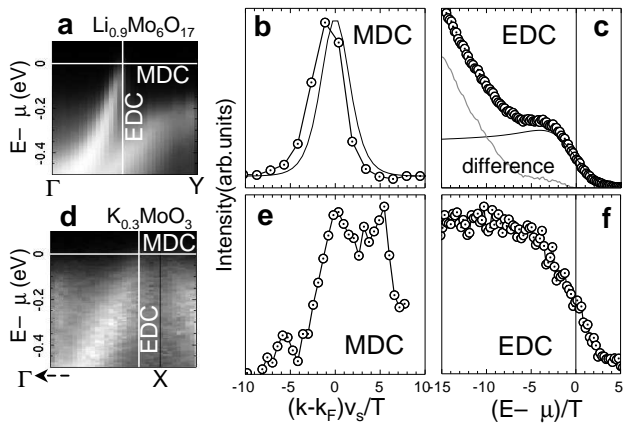}}
    \par}

\caption{\label{fig:mdc-edc}
  ARPES intensity map for the Li purple bronze (a), and its MDC (b) and
  EDC (c) cuts, and similar plots for the K blue bronze (d, e, and f).  In
  (b) and (c), theory curves of the TL model, as used in Fig.\ 
  \ref{fig:li}, are shown as thin solid lines.  In (c), the difference between
  the data and the theory is shown as a gray solid line, representing the
  non-$\mu$-crossing band.  The K blue bronze has three crossings showing
  as three peaks in the MDC, and here we focus on the central crossing,
  but others lead to the same conclusion.}
\end{figure}

\textbf{4. Sharp MDC, Broad EDC}\ \ Orgad et al.\ \cite{orgad-efrac} noted
that the TL spectral functions show a sharp momentum distribution curve
(MDC) at $\mu$ and a broad energy distribution curve (EDC) at
\textbf{k}$_{\text{F}}$ while for the FL both are sharp.   The
mechanism for this contrast in the theory is the generic 
1-d kinematic restrictions on momentum, but not on energy, 
for electron fractionalization into one or
more kinds of dispersive density waves, as happens in, but is not 
restricted to, the TL (or LE) model.  They reported
this striking MDC/EDC contrast for cuprates, especially the static stripe
system La$_{1.25}$Nd$_{0.6}$Sr$_{0.15}$CuO$_4$, but also the Bi2212. 
As seen in Fig.\ \ref{fig:mdc-edc}, our data
for quasi-1-d metals indeed show this contrast, a small momentum offset
from the theory for the Li purple bronze \cite{mdc-offset}
notwithstanding.  So does the Na purple bronze data of 
Fig.\ \ref{fig:napb2212} (also see \cite{gweon-jpcm}) once we recognize the
\textbf{k}-independent (badly scattered charge wave) weight as part of the
intrinsic spectrum.  It is precisely the \textbf{k}-independent weight
in the Bi2212 spectra that enabled the identification made by
Orgad et al.\ \cite{orgad-efrac} of this signature in Bi2212 and 
our discussion here enables us to offer an explicit 
proposal as to the origin of the weight within a fractionalization picture.  
Strong arguments that this weight is intrinsic to the spectral function
have been given previously in Refs.\ \cite{sawatzky} and \cite{liu}.

Note that this MDC/EDC contrast requires at least
moderately large $\alpha$ \cite{orgad-efrac}, further supporting 
our conclusion of a non-zero but small $\alpha$
for Bi2212 from the combined analysis of signatures 1 and 2.
Note also that $\text{v}_{\text{s}}$ is used to scale the
momentum axis of the MDC plots.  While it means the spin wave velocity for
the TL model (and therefore for the Li purple bronze), its generic meaning
is the smallest velocity of density waves involved, and for the poorly
understood spin-gapped K blue bronze we use the value
of 1eV{\AA}  to make its MDC width similar to that of
the Li purple bronze.  

\section{Summary}

To summarize, our ``score card'' representation of the electron
fractionalization signatures is the following: Li purple bronze (1,2,3,4),
K blue bronze (1,2,4), Na purple bronze (1,2,3,4), Bi2212 (2,3,4).  Our
electron fractionalization signatures are ubiquitous in the sense that
each of the examined materials shows at least three signatures.  The Li 
purple bronze displays all four and as such is a quasi-1-d LL paradigm.  
The missing signature in 
the quasi-1-d K-blue bronze can be attributed to complexities of a
spin gap and its incipient CDW ordering.  Neither of these materials
seems to require the ideas of Ref.\ \cite{baer-prl}.  It is 
especially notable that the last two of our materials are quasi-2-d 
and that the Na purple bronze displays all four signatures within
the framework of our ``melted holon'' interpretation.  Like
the Na purple bronze, Bi2212 could well be fractionalized with 
nonzero $\alpha < 0.5$ and badly scattered charge waves, but
with either $\alpha$ or the energy range of power law behavior too small 
to detect in the $\mu$ onset of its angle integrated spectrum.  We note that
the \textbf{k}-independent spectral weight of both materials demands some
explanation outside of the FL picture.  We argue here that the 
Na purple bronze is
a kind of Rosetta stone that enables the recognition of the effects
of disorder on fractionalization in the Bi2212 spectrum.

Overall, it is remarkable that all these systems display common, 
generic signatures within our generalized electron
fractionalization scheme regardless of their different, and often not fully
understood, low $T$ physics, and different global dimensionalities.  Our
findings are a strong hint of a bigger picture in which fractionalization
plays a central role.  In closing we mention that another important
general fractionalization signature that is implicit in our use 
of the theory of Ref.\ \cite{orgad-finite-T-TL-LE} is that 
of quantum critical scaling in the energy and temperature dependence 
of ARPES lineshapes.  Such scaling has already been reported 
for Bi2212 lineshapes \cite{qcarpes} and our plans for the future include
such studies for the other materials we have discussed.

\begin{acknowledgments}

  GHG and JWA acknowledge useful discussions with S.A. Kivelson. For the
  data of Fig.\ \ref{fig:ar}(d), we are indebted to J. Marcus and C.
  Schlenker, to W.P. Ellis, R. Claessen and F. Reinert, and to Z.-X. Shen
  for providing samples, for participation in the experiments, and for use
  of his end-station, respectively. This work was supported by the U.S.
  NSF grant No.\ DMR-99-71611 and the U.S. DoE contract No.\ 
  DE-FG02-90ER45416 at U.  Mich.

\end{acknowledgments}

$^\dagger$ Current address: Lawrence Berkeley National Laboratory,
MS 7-100, 1 Cyclotron Road, Berkeley, CA 94720


\begin{thebibliography}{99}
  
\bibitem{landau-FL} L.D. Landau, Sov.\ Phys.\ JETP \textbf{30}, 1058
  (1956).
  
\bibitem{haldane-LL} F.D.M. Haldane, J. Phys.\ C \textbf{14}, 2585 (1981).
  

\bibitem{TLmodel-T} S. Tomonaga, Prog.\ Theor.\ Phys.\ \textbf{5}, 544 (1950).
  

\bibitem{TLmodel-L} J. M. Luttinger, J. Math Phys.\ \textbf{4}, 1154 (1963).
  

\bibitem{anderson-efrac} P.W. Anderson, Physics Today \textbf{50}, 42
  (1997).
  
\bibitem{kivelson-efrac} S.A. Kivelson, Synthetic Metals \textbf{125}, 99
  (2002).
  
\bibitem{orgad-efrac} D. Orgad et al., Phys.\ Rev.\ Lett.\ \textbf{86},
  4362 (2001).

\bibitem{LEmodel} A. Luther and V. J. Emery, Phys.\ Rev.\ Lett.\ \textbf{33},
  589 (1974).


\bibitem{dardel-prl} B. Dardel et al., Phys.\ Rev.\ Lett.\ \textbf{67},
  3144 (1991).
  
\bibitem{grioni-voit} M.\ Grioni and J.\ Voit, in {\em Electron
    spectroscopies applied to low-dimensional materials} (eds.\ H.P.
  Hughes and H.I. Starnberg) p.\ 209 (Kluwer, Dordrecht, 2000).
  
\bibitem{gweon-jesrp} G.-H. Gweon et al., J. Elec.\ Spectro.\ Rel.\ 
  Phenom.\ \textbf{117-118}, 481 (2001).



\bibitem{denlinger-Li-purple-prl} J.D. Denlinger et al., Phys.\ Rev.\ 
  Lett.\ \textbf{82}, 2540 (1999).

 \bibitem{gweon-prlComm} G.-H. Gweon et al., Phys.\ Rev.\ Lett.\
  \textbf{85}, 3985 (2000).
 
  
\bibitem{gweon-physicaB} G.-H. Gweon et al., Physica
  B, in print.

\bibitem{orgad-finite-T-TL-LE} D. Orgad, Philos.\ Mag.\ B \textbf{81}, 377
  (2001).

 
\bibitem{pouget-Li-struct} J.P. Pouget, private commmunication.
  

\bibitem{degiorgi-purple-optics} L. Degiorgi et al., Phys. Rev. B
 \textbf{38}, 5821 (1988).

\bibitem{greenblatt} M. Greenblatt et al., Solid State Commun.\
  \textbf{51}, 671 (1984).
  

\bibitem{schlenker-Mo-oxide-book} C. Schlenker, ed. {\em Low-Dimensional
    Electronic Properties of Molybdenum Bronzes and Oxides} (Kluwer
    Academic Publishers, Bordrecht/Boston/London, 1989). 

\bibitem{whangbo} M. H. Whangbo et al., Science \textbf{252}, 96 (1991).

\bibitem{gweon-hidden-1d-FS} G.-H. Gweon et al., Phys.\ Rev.\ B
  \textbf{55}, R13353 (1997).
  
  
\bibitem{kaminski-prl} A. Kaminski et al., Phys.\ Rev.\ Lett.\ 
  \textbf{86}, 1070 (2001).

  \bibitem{dessau-Bi2212-prl93} D.S. Dessau et al., Phys.\ Rev.\ Lett.\
  \textbf{71}, 2781 (1993).

\bibitem{Carlson} E. W. Carlson et al., Phys.\ Rev.\ B 
  \textbf{62}, 3422 (2000).

\bibitem{mckenzie-SRO-non-FL} R.H. McKenzie and D. Scarratt, Phys.\ Rev.\ 
  B \textbf{54} R12709 (1996).

  
\bibitem{ssrl-data} The data in Fig.\ \ref{fig:ar}(d) were taken at 
  Beamline 5 of the Stanford Synchrotron Radiation Laboratory, with $h\nu
  = 22.4$ eV, $\Delta E = 43$ meV and $\Delta \theta = 2^o$.

\bibitem{claessen-prl-TiTe2} R. Claessen et al. Phys.\ Rev.\ Lett.\ 
  \textbf{69}, 808 (1992).
  

\bibitem{claessen-prb-TiTe2} R. Claessen et al. Phys.\ Rev.\ B 
  \textbf{54}, 2453 (1996).


\bibitem{perfetti-TiTe2} L. Perfetti et al. Phys.\ Rev.\ B 
  \textbf{64}, 115102 (2001).

 
 
  
\bibitem{alpha-anisotropy} The data near the 2-d BZ edge (i.e.\ near the
  ($\pi$,0) point) [A. Kaminski et al., private communication.]  yield
  $\alpha$=0.1, signaling some \textbf{k} dependence of $\alpha$.  The
  small band dispersion near ($\pi$,0) precludes an AR analysis.
  
  
  
\bibitem{penc-1d-Hubbard} K. Penc, F. Mila, and H. Shiba, Phys.\ Rev.\ 
  Lett.\ \textbf{75}, 894 (1995).
  
\bibitem{johnston} D.C. Johnston, Phys.\ Rev.\ Lett.\ 52, 2049 (1984).

\bibitem{anderson-ren-line-shape} P.W. Anderson and Y. Ren, in {\em High
    Temperature Superconductivity} (eds.\ K. S. Bedell et al.) p.3
  (Addison-Wesley, Redwood City, 1990).

\bibitem{AR disordered}  P.W. Anderson has suggested to us that
the AR lineshape might obtain for the "melted holon" scenario 
discussed for signature number 3.
  
\bibitem{valla-Mo} T. Valla et al., Phys.\ Rev.\ Lett.\ \textbf{83}, 2085 (1999).

  
\bibitem{meden-voit-LL-line-shape} V. Meden and K. Sch\"{o}nhammer, Phys.\ 
  Rev.\ B \textbf{46}, 15753 (1992); J. Voit, Phys.\ Rev.\ B \textbf{47},
  6740 (1993).
  
\bibitem{gweon-jpcm} G.-H. Gweon et al., J. Phys.\ Condens.\ Matter
  \textbf{8}, 9923 (1996).

\bibitem{anderson-private} P.W. Anderson, private communication, 1995.
  
\bibitem{balents-coupled-LL} L. Balents and M. P. A. Fisher, Phys.\ Rev.\ 
  B \textbf{53}, 12133 (1996).
  
\bibitem{mdc-offset} While we do not have a firm explanation for this
  small effect, a slightly larger experimental MDC width and the matrix
  element modulation that gives a strong suppression of weights for
  \textbf{k} $>$ \textbf{k}$_{\text{F}}$ would be a likely explanation.

\bibitem{sawatzky} G.A. Sawatzky, in  {\em High
    Temperature Superconductivity} (eds.\ K. S. Bedell et al.) p.297
  (Addison-Wesley, Redwood City, 1990).


\bibitem{liu} L. Z. Liu, R. O. Anderson and J.W. Allen, 
J. Phys.\ Chem.\ Sol.\
  \textbf{52}, 1473 (1991)

\bibitem{baer-prl} P. Starowicz et al., Phys.\ Rev.\ Lett.\ 
\textbf{89}, 256402 (2002).


\bibitem{qcarpes} T. Valla et al., Science \textbf{285}, 2110
 (1999).

  
\end{thebibliography}
\end{document}